# Unconventional critical behaviour in weak ferromagnets $Fe_{2-x}Mn_xCrAl$ ($0 \leq x < 1$)

Kavita Yadav, Dheeraj Ranaut, and K. Mukherjee

School of Basic Sciences, Indian Institute of Technology, Mandi, Himachal Pradesh-175005, India

**Abstract** – Recent studies on weak ferromagnets $Fe_{2-x}Mn_xCrAl$ ($0 \leq x < 1$) reveal the presence of cluster glass (CG) and Griffiths phase (GP) below and above the ferromagnetic transition temperature ($T_C$) [(2019) *Sci. Rep.* **9** 15888]. In this work, the influence of these inhomogeneous phases on the critical behaviour (around $T_C$) of the above-mentioned series of alloys has been investigated in detail. For the parent alloy $Fe_2CrAl$, γ is estimated as ~ 1.34, which lies near to ordered 3D Heisenberg model, whereas obtained value of β ~ 0.273 does not belong to any universality class. With increment in Mn concentration, both exponents γ and β increases, where γ and β approaches disordered 3D Heisenberg model and ordered 3D Heisenberg model, respectively. The observed deviation in γ and unconventional value of δ can be ascribed to the increment in GP with Mn-concentration. The trend noted for β can be attributed to the increment in CG regime with an increase in Mn-content. The self-consistency and reliability of the estimated exponents are verified by the Widom scaling relation and scaling equations of state. Our studies indicate that the critical phenomenon of $Fe_{2-x}Mn_xCrAl$ ($0 \leq x < 1$) alloys possibly belong to a separate class and is not described within the framework of any existing universal model.

**Introduction.** - In the past few decades, Heusler alloys are extensively investigated due to interesting diverse magnetic phenomenon like localized and itinerant magnetism, antiferromagnetism, helimagnetism, Pauli paramagnetism and heavy fermionic behaviour [1-4] exhibited by them. Interestingly, majority of the Heusler alloys order ferromagnetically and saturate in weak field [5]. This long-range ordering gets significantly modified by various substitutions, anti-site disorder, and variations in stoichiometry. For example, there is an evolution of ferromagnetic (FM) behaviour in $Fe_2V_{1-x}Cr_xAl$ on substitution of Cr at V site [6]. This substitution alters the FM coupling between the clusters. In $Fe_2Cr_{1-x}Mo_xAl$ series of alloys, with Mo substitution, a significant decrement in value of $T_C$ is noted [7]. In this context, analysis of critical exponent is useful to understand the role of a substitution, structural disorder, or stoichiometric variations in modifying the FM interactions. This method has been used widely by researchers to investigate the second order phase transition (SOPT) in Heusler alloys [7-12]. For instance, Phan et al. [8] showed that Sn substitution at Mn site in $Ni_{50}Mn_{50-x}Sn_x$ affects the short-range FM interactions and led to cross over from short range to long range FM ordering. In a similar way, the role of various substitutions is studied through critical exponent analysis as reported in $Ni_{47}Mn_{40}Sn_{13-x}Cu_x$, $Ni_{43}Mn_{46}Sn_8X_3$ (X=In and Cr), $Ni_{2.2}Mn_{0.72-x}V_xGa_{1.08}$ [10-12]. Additionally, disorder also influences the critical phenomenon around FM transition [9, 13, 14, and 15]. In $Pr_{0.5}Sr_{0.5-x}Ag_xMnO_3$, it is noted that the obtained critical exponents do not belong to a single universality class. The increment in Ag concentration leads to an augmentation in anti-site disorder which results in the short range interaction in the system [15]. In $Ni_{50}Mn_{37}Sn_{13}$ it is observed that Gd substitution at Ni site led the system from short range FM order to long range FM order [9]. Presence of disorder in transition metal-based oxide systems can lead to the formation of Griffiths phase (GP). This phase affects the long-range ordering in these systems and unconventional critical exponents are reported [16, 17]. However, the effect of GP on critical exponents is still poorly understood in the Heusler alloys.

Recently, we have investigated the structural and magnetic properties of $Fe_{2-x}Mn_xCrAl$ ($0 \leq x \leq 1$) Heusler alloys [18]. Our results reveal that parent alloy $Fe_2CrAl$ undergoes from FM to paramagnetic (PM) transition near $T_C \sim 202$ K. It exhibits cluster glass (CG) phase below $T_{f1} \sim 3.9$ K and GP above $T^* \sim 300$ K. It is observed that with Mn substitution (as $x \rightarrow 1$) $T_C$ shifts significantly towards lower temperature, with complete disappearance of FM behaviour in FeMnCrAl. Additionally, the alloys show CG-like phase at low temperatures. Also, GP is found to be persistent in all alloys, with a decrement in $T^*$ with Mn concentration. Hence it is of interest to investigate: (a) how increment of Mn substitution at Fe site in $Fe_2CrAl$ and presence of CG phase influences the critical phenomenon and FM interactions in the vicinity of $T_C$? and (2) whether the presence of GP is always a precursor to the observed deviation of critical exponents values from the values noted in universality class? Hence, to shed some light on the above-mentioned issues, in this letter, we have investigated the critical behaviour of $Fe_{2-x}Mn_xCrAl$ ($0 \leq x < 1$) Heusler alloys in the vicinity of $T_C$.

**Experimental.** - The alloys, $Fe_2CrAl$, $Fe_{1.75}Mn_{0.25}CrAl$, $Fe_{1.5}Mn_{0.5}CrAl$, and $Fe_{1.25}Mn_{0.75}CrAl$ are the same as those reported in Ref [18]. Structural characterization of these alloys has been already reported in Ref. 18. From this study, we have concluded the presence of anti-site disorder in all studied alloys. It is also noted that anti-site disorder increases with the Mn content. Additionally, in Ref. 19, morphological and compositional analysis of all these alloys has been carried out. It confirms the homogenous distribution of all elements in corresponding alloys. Also, it indicates that all alloys under investigation are homogenously disordered. Magnetic field (*H*) dependent magnetization (*M*) measurements have been carried using Magnetic Property Measurement System (MPMS), Quantum Design, U.S.A. Rectangular shaped samples are used to obtain the *M-H* isotherms. The isotherms are collected in a close temperature interval (~ 1 K). Each isotherm is measured after cooling the sample from room temperature (after the removal of remanent magnetic field) to the measurement temperature. For proper stabilization of temperature, 10 minutes wait time is given before recording each isotherm.

**Results.** - In order to analyze the critical phenomenon in the vicinity of the $T_C$, where a magnetic system undergoes a SOPT from PM phase to FM state, a set of critical exponents are determined. According to the scaling hypothesis, the spontaneous magnetization $M_s$ (*T*) below $T_C$, initial susceptibility $\chi_0^{-1}$ (*T*) above $T_C$ and magnetization *M* at $T_C$ shows following power law dependence for SOPT [20]:

$$M_S(T) = M_0(-\varepsilon)^\beta, \varepsilon < 0 \ldots\ldots\ldots\ldots (1)$$

$$\chi_0^{-1}(T) = \Gamma(-\varepsilon)^\gamma, \varepsilon > 0 \ldots\ldots\ldots\ldots (2)$$

$$M = XH^{\frac{1}{\delta}}, \varepsilon = 0 \ldots\ldots\ldots\ldots\ldots\ldots (3)$$

where $M_0$, $\Gamma$ and X are the critical amplitudes; $\beta$, $\gamma$, and $\delta$ are the critical exponents; and $\varepsilon = (T-T_C)/(T_C)$ is the reduced temperature. The magnetic equation of the state is a relationship among the variables *M* (*H*, $\varepsilon$), *H*, and *T*. Using the scaling hypothesis this can be expressed as

$$M(H,\varepsilon) = \varepsilon^\beta f_\pm(\frac{H}{\varepsilon^{\beta+\gamma}}) \ldots\ldots\ldots\ldots\ldots (4)$$

where $f_+(T>T_C)$ and $f_-(T<T_C)$ are the regular functions. In terms of renormalized magnetization m = $\varepsilon^{-\beta}M(H,\varepsilon)$ and h = $\varepsilon^{-(\beta+\gamma)}H$, the eqn. (4) can be written as

$$m = f_\pm(h) \ldots\ldots\ldots\ldots\ldots\ldots\ldots\ldots (5)$$

The above equation implies that for true scaling relations and right choice of $\beta$, $\gamma$, and $\delta$ values; scaled *m* plotted as a function of scaled *h* will fall on two universal curves: one above $T_C$ and another below $T_C$. This is an important criterion for validity of critical regime. Generally, exponents in the asymptotic regime ($\varepsilon \rightarrow 0$) show universal properties. However, exponents often show various systematic trends or crossover phenomena as one approaches $T_C$ [21, 22]. This occurs when there is presence of disorder or couplings in the system. Due to this reason, the temperature dependent effective exponents for $\varepsilon\neq0$ are introduced. These effective exponents are non-universal properties, and are defined as

$$\beta_{eff} = \frac{d[lnM_S(\varepsilon)]}{d(ln\varepsilon)}, \gamma_{eff} = \frac{d[ln\chi_0^{-1}(\varepsilon)]}{d(ln\varepsilon)} \dots\dots\dots (6)$$

In the asymptotic limit, these exponents should approach universal properties.

Conventionally, Arrott plots are used to analyse the critical region around $T_C$. In this method, the magnetic isotherms are plotted in the form of $M^2$ Vs $H/M$, which constitute a set of parallel straight lines around $T_C$ [23]. This plot follows mean field theory ($\beta$ = 0.5, $\gamma$ = 1) and isotherms exhibit linear behaviour in high field region. It also provides us the magnitude of the $\chi^{-1}_0$ (T) and $M_s$ (T) as an intercept on $H/M$ and $M^2$ axis, respectively. Arrott plot for all the alloys is plotted around $T_C$ (shown in supplementary material Fig. S1 (a)-(d)). For all the alloys, it is observed that all the curves in the plot show non-linear behaviour having downward curvature even in the high field regime. This indicates that the critical behaviour of $Fe_{2-x}Mn_xCrAl$ (0≤x<1) alloys cannot be described based on the mean field theory. Moreover, according to Banerjee criterion the downward curvature indicates the second order nature of the phase transition [24]. A generalized form of this analysis, known as Modified Arrott plot (MAP) involves plotting $M^{1/\beta}$ vs $H/M^{1/\gamma}$ in the critical regime. It is given by the following equation of state:

$$\left(\frac{H}{M}\right)^{\frac{1}{\gamma}} = a\left(\frac{T-T_C}{T}\right) + bM^{\frac{1}{\beta}} \dots\dots\dots\dots (7)$$

where *a* and *b* are the constants. However, determination of critical exponents through this method is non-trivial task as $\beta$ and $\gamma$ are two free parameters are involved in eqn. 7. This can lead to systematic errors in the obtained value of exponents. Hence, for the proper selection of $\beta$ and $\gamma$ an iterative method has been followed. In this method, initial values of critical exponents are taken as $\beta$ = 0.365 and $\gamma$ = 1.386 (same as theoretical 3D Heisenberg model). These values are substituted in eqn. 7 to generate a MAP. Figure 1 (a) - (d) shows the MAP for $Fe_2CrAl$, $Fe_{1.75}Mn_{0.25}CrAl$, $Fe_{1.5}Mn_{0.5}CrAl$ and $Fe_{1.25}Mn_{0.75}CrAl$ at different temperatures. From the linear extrapolation of isotherm, the value of $(M_S)^{1/\beta}$ and $(\chi_0^{-1})^{1/\gamma}$ as an intercept on $M^{1/\beta}$ and $H/M^{1/\gamma}$ axis are obtained, respectively. These values of $M_S(T)$ and $\chi_0^{-1}(T)$ are used to fit in eqn. (1) and eqn. (2), respectively. According to these equations, the slope of straight-line fitting of log($M_S$) vs log ($\varepsilon$) and log($\chi_0^{-1}$) vs log($\varepsilon$) gives us the new value of $\beta$ and $\gamma$, respectively. It is important to note that that during the straight-line fitting, the free parameter $T_C$ is adjusted in eqn. (1) and eqn. (2) such that a best fit can be obtained. The new values of $\beta$ and $\gamma$ are again used to construct a new MAP. This process is continued until stable values of $\beta$, $\gamma$ and $T_C$ are obtained (as listed in table 1). Using this method, for each alloy, a set of parallel isotherms has been generated. The ultimately obtained values of $M_S(T)$ and $\chi_0^{-1}(T)$ are then used to again estimate the values of critical exponents and $T_C$ through scaling law. Here, for each alloy, $M_S(T)$ and $\chi_0^{-1}(T)$ are plotted as a function of temperature as shown in Figure 2 (a)-(d). Using these values of $M_S(T)$ and $\chi_0^{-1}(T)$, the obtained values of β, γ and $T_C$ are listed in table 1. As noted from table 1, the estimated values from both techniques (scaling law and MAP) match well with each other.

For more accurate determination of critical exponents as well as $T_C$, $M_S(T)$ and $\chi_0^{-1}(T)$, the data is analysed using Kouvel-Fischer (KF) plot [25]. In this technique, $M_S (dM_S/dT)^{-1}$ vs $T$ and $\chi_0^{-1}(d\chi_0^{-1}/dT)^{-1}$ vs $T$ yield straight lines with slopes $1/\beta$ and $1/\gamma$, respectively. The advantage of this plot is that prior knowledge of $T_C$ is not required as the intercept of the fitted straight lines yields $T_C$. KF plots for all alloys has been presented in Figure 3 (a)-(d). The estimated exponents and $T_C$ are listed in table 1. Interestingly, the values of critical exponents and $T_C$ estimated through KF plot and MAP matches reasonably well.

Critical isotherm $M(H, T_C)$ vs $H$ for each alloy is plotted (shown in supplementary material Fig. S2 (a)-(d)) and insets represent the same plot in log-log scale). According to eqn. 3, log $M$ vs log $H$ plot will give a straight line with slope $1/\delta$. From the straight-line fitting, the values of δ are determined. Also, δ has been calculated from the Widom-scaling relation [26] which is

$$\delta = 1 + \frac{\gamma}{\beta} \quad \text{...................} \quad (8)$$

using the value of γ and β obtained from MAP studies, the obtained values of δ are 5.69±0.08, 5.94±0.05, 5.28±0.08 and 5.56±0.01 for $Fe_2CrAl$, $Fe_{1.75}Mn_{0.25}CrAl$, $Fe_{1.5}Mn_{0.5}CrAl$ and $Fe_{1.25}Mn_{0.75}CrAl$, respectively. These values are very close to the value obtained from the critical isotherms. Hence, in the present study, the estimated critical exponents are accurate and self-consistent. All the critical exponents estimated from different methods along with the theoretical values are given in table 1.

The experimentally obtained values of critical exponents do not match with any of conventional universality classes. Hence, to check whether the obtained parameters can generate the scaling equation of state (eqn. 5), the scaled $m$ is plotted as a function of scaled $h$ for each alloy as shown in Figure 4 (a)-(d). Inset of the Fig. 4 (a)-(d) shows the log-log scale of the same plot. It can be clearly noted that the scaling law is satisfied in each case i.e. all the obtained isotherms diverge into two separate curves: one below $T_C$ and one above $T_C$. Furthermore, the reliability of the critical exponents and $T_C$ has been re-checked using more rigorous method where $m^2$ is plotted as function of $h/m$ [27], given in supplementary information Fig. S3 (a)-(d)). From Fig. S3 (a)-(d), for each alloy, it is observed that the data fall into two curves: one above $T_C$ and one below $T_C$.

As the critical exponents obtained from various methods do not fall in any universality class, it is important to determine whether the values of γ and β matches with any universality class under asymptotic limit. For this purpose, effective critical exponents are determined as a function of reduced temperature(ε). It can be observed from the Fig. 5(a)-(d) that both parameters exhibit non-monotonic change with variation in ε. In case of $Fe_2CrAl$, it can also be seen that $β_{eff}$ and $γ_{eff}$ show slight dip (at ε=-0.02) and peak (at ε=0.05) before approaching asymptotic limit (ε→0). Similar trend in $β_{eff}$ and $γ_{eff}$ for other compositions is observed. Here, the values of $β_{eff}$ and $γ_{eff}$ at $ε_{min}$ do not match with any predicted universality class. Additionally, the data do not fully collapse into two separate branches with values of $β_{eff}$ and $γ_{eff}$ estimated at $ε_{min}$. This can arise due to the following reasons: (i) $ε_{min}$ does not lie in the asymptotic region and $T_C$ must be approached more closely for asymptotic exponents or (ii) $ε_{min}$ lies in asymptotic region as similar type of disagreement of effective critical exponents (with any universality classes) is also noted for other disordered materials [28]. In case of crystalline FM, $γ_{eff}(ε)$ shows a monotonous decrement with an increment in ε, whereas a peak is observed in amorphous FM [27]. From Fig. 5 (a)-(d), it is observed that the temperature variation of the effective critical exponents is similar to behaviour seen in disordered FM. Thus, the above observations signify the influence of disorder on the critical exponent's values. In our case, there is presence of anti-site disorder between Fe and Al [18] which increases with Mn concentration. This disorder results in the formation of inhomogeneous magnetic phases, which is responsible for the observed unconventional values of critical exponents in $Fe_{2-x}Mn_xCrAl$.

**Discussions. -** The obtained values of critical exponents of $Fe_{2-x}Mn_xCrAl$ are unconventional and do not belong to any universality class. In our previous studies, it is reported that $Fe_2CrAl$ undergoes FM to PM transition near $T_C$ ~ 202 K with the presence of GP above $T_C$. With increment in Mn concentration, $T_C$ decreases, and the temperature regime between FM and GP increases as shown in the phase diagram [18]. In the present study, we have noted that the values of $T_C$ (corresponding to each alloy) obtained through various techniques matches reasonably well with the previous reported values [18]. The value of γ for the parent alloy is found to be 1.34 which is near to that reported for an ordered 3D Heisenberg model. This value increases with increasing Mn-concentration and is found to be 1.6 for $Fe_{1.25}Mn_{0.75}CrAl$. It is similar to that reported for a disordered 3D Heisenberg model. Physically, γ represents the degree of divergence of χ(T) at $T_C$, smaller the value of γ, sharper will be the divergence. For the parent alloy, γ is smaller as compared to $Fe_{1.25}Mn_{0.75}CrAl$, which is in accordance to the observation of the sharp transition in the former case. Also, the larger magnitude of γ indicates the broader temperature range of PM-FM transition. The observed trend in the value of γ is consistent with the increment of temperature regime of GP. The Yang-Lee theory [16] of phase transition predicts that the singularity of the GP can lead to unusual critical behaviour i.e. a discontinuity in the $M$ ($H$) at $T=T_C$. It is reflected in observed larger values of critical exponent δ. In the present case, unusual larger values of δ are noted for these alloys. This behaviour suggests that the GP affects γ and δ. As reported in Ref 18, in these systems, GP arises due to anti-site disorder. Interestingly, the observed non-monotonic temperature dependent behaviour of $\gamma_{eff}$ also reflects the influence of disorder on the critical exponents. Here, it can be speculated that the random anti-site disorder can cause a strong broadening of the distribution of the local exchange fields owing to competition between AFM and FM exchange interactions. Similar behaviour was also noted in $Fe_{100-x}Pt_x$ alloy [29]. In these alloys, the value of critical exponent γ was enhanced by increasing the metallurgical site disorder.

In $Fe_2CrAl$, it is observed that there is a presence of CG regime in low temperature regime (below $T_C$). This regime increases with Mn-substitution. For $Fe_2CrAl$, the value of β is 0.273, which does not belong to any universality class. With increasing Mn content, the value of β increases and approaches ordered 3D Heisenberg model, as found for $Fe_{1.25}Mn_{0.75}CrAl$ (β=0.347). Physically, β represents the growth of spontaneous magnetization below $T_C$ i.e. smaller value indicates faster growth. In the present case, value of β is smaller for parent alloy as compared to $Fe_{1.25}Mn_{0.75}CrAl$, implying that the growth of $M_S$ is faster in the former alloy. With Mn substitution, the rate of growth decreases near $T_C$, which is a consequence of increased CG phase region. However, the obtained values of β (= 0.273) for $Fe_2CrAl$ in our case does not match well with the earlier reported value of β (= 0.42) for the same alloy [30]. This discrepancy in the obtained value of β can arise due to presence of short-range correlations (in CG phase) below $T_C$ in our case and has not been reported in the latter case.

Hence, it can be said that critical exponents for disordered ferromagnetic systems is not in accordance with any conventional universality classes. Both γ and β are affected due to the presence of GP and CG phase, respectively. Unconventional behaviour of critical exponents is not unusual and has also been observed in various alloys as well as oxides. For example, large value of β (=0.43) is found due to phase segregation in $La_{1-x}Sr_xCoO_3$ compound [31]. Similarly, $Gd_{80}Au_{20}$ exhibits unconventional exponents β=0.44 and γ=1.29, which arises due to spin dilution on non-magnetic ion substitution [32]. Interestingly, due to presence of GP in $La_{0.79}Ca_{0.21}MnO_3$, larger values of γ and δ are observed [16]. Thus, our results suggest that critical phenomenon in $Fe_{2-x}Mn_xCrAl$ cannot be explained based on the existing universality classes.

**Summary. -** The influence of CG and GP on the critical phenomenon near PM-FM phase transition of $Fe_{2-x}Mn_xCrAl$ (0≤x<1) has been investigated. This transition is identified to be second order in nature. The critical exponents γ, β, and δ estimated from various techniques match reasonably well. For $Fe_2CrAl$, the estimated value of β is smaller than ordered 3D Heisenberg model whereas, γ is found to be near to this model. Along the series, both exponents' β and γ show an increasing trend. For all alloys, the temperature dependence of $γ_{eff}$ and $β_{eff}$ resemble to disordered ferromagnets, signifying the effect of anti-site disorder. This disorder induces Griffiths like properties above $T_C$, which is reflected in the unconventionally larger values of γ and δ. Additionally, the observed trend in β can be attributed to increment in CG regime due to Mn-substitution. Our study will be helpful to understand the effect of inhomogeneous magnetic phase (below and above $T_C$) on the critical behaviour of weak ferromagnetic Heusler alloys.

***

K.M. acknowledges the financial support from a research grant (Grant No. 03(1381)/16/EMR-II) from SERB, India. The authors acknowledge IIT Mandi for providing the experimental facilities.

**Table 1** Values of the exponents β, γ and δ as determined from the modified Arrott plot, Kouvel-Fischer plot and critical isotherm for $Fe_2CrAl$, $Fe_{1.75}Mn_{0.25}CrAl$, $Fe_{1.5}Mn_{0.5}CrAl$ and $Fe_{1.25}Mn_{0.75}CrAl$. The theoretically predicted values for various universality classes are also listed for comparison.

| Composition | Ref. | Method | β | $T_C (M_S)$ K | γ | $T_C(\chi_0)$ K | δ | $T_C$ [17] |
|---|---|---|---|---|---|---|---|---|
| $Fe_2CrAl$ | This work | MAP | 0.2730 | 201.89±0.37 | 1.34 | 201.11±0.14 | 5.91 | 202 K |
| | | KF plot | 0.273±0.01 | 201.06±0.04 | 1.35±0.02 | 201.01±0.03 | 5.94±0.03 | |
| | | CI | - | - | - | - | 5.69±0.08 | |
| $Fe_{1.75}Mn_{0.25}CrAl$ | This work | MAP | 0.2822 | 120.90±0.05 | 1.42 | 121.19±0.06 | 6.03 | 120 K |
| | | KF plot | 0.282±0.02 | 120.34±0.09 | 1.41±0.03 | 121.02±0.02 | 6±0.01 | |
| | | CI | - | - | - | - | 5.94±0.05 | |
| $Fe_{1.5}Mn_{0.5}CrAl$ | This work | MAP | 0.3281 | 48.6±0.25 | 1.46 | 47.32±0.05 | 5.45 | 48 K |
| | | KF plot | 0.326±0.02 | 46.67±0.01 | 1.47±0.05 | 47.41±0.01 | 5.45±0.02 | |
| | | CI | - | - | - | - | 5.28±0.08 | |
| $Fe_{1.25}Mn_{0.75}CrAl$ | This work | MAP | 0.3474 | 26.86±0.09 | 1.6 | 27.15±0.05 | 5.61 | 27 K |
| | | KF plot | 0.345±0.13 | 26.79±0.06 | 1.61±0.04 | 27.10±0.04 | 5.60±0.07 | |
| | | CI | - | - | - | - | 5.56±0.01 | |
| **Mean field theory** | 25 | Theory | 0.5 | | 1 | | 3 | |
| **3D Heisenberg model (O*)** | 25 | Theory | 0.365 | | 1.386 | | 4.80 | |
| **3D Heisenberg model (D*)** | 29 | Theory | 0.5 | | 2 | | 5 | |
| **3D Ising model** | 25 | Theory | 0.325 | | 1.241 | | 4.82 | |
| **3D XY model** | 25 | Theory | 0.346 | | 1.316 | | 4.81 | |

*O= ordered D= disordered

**Figures-**

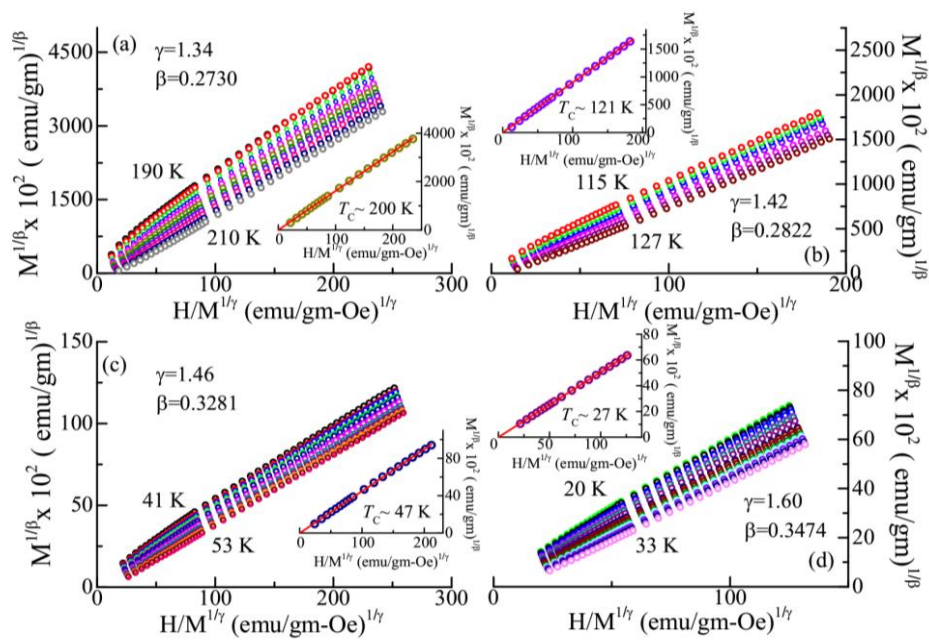

Figure 1 Modified Arrott plot ($M^{1/\beta}$ vs $H/M^{1/\gamma}$) of (a) $Fe_2CrAl$ (b) $Fe_{1.75}Mn_{0.25}CrAl$ (c) $Fe_{1.5}Mn_{0.5}CrAl$ and (d) $Fe_{1.25}Mn_{0.75}CrAl$ with estimated critical exponents (as also listed in table 1). Insets: MAP of each alloy at $T_C$. Red straight line represents the linear fit of isotherm at $T_C$. Only few isotherms are presented for clarity.

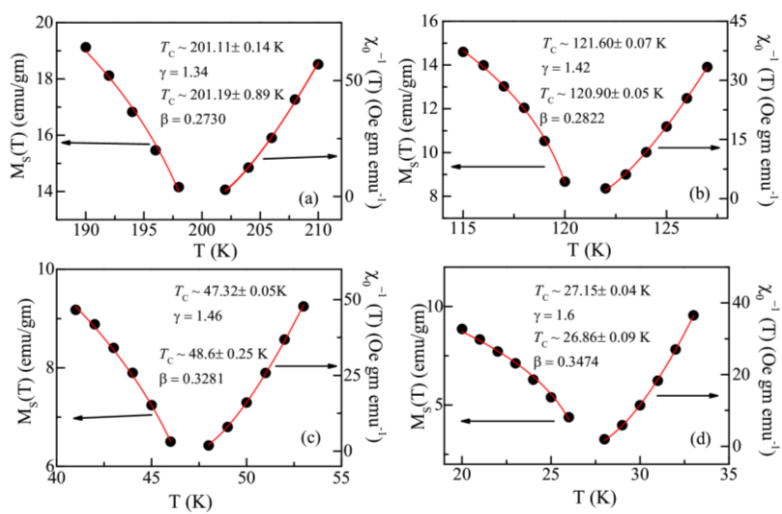

Figure 2 Temperature dependent behaviour of spontaneous magnetization ($M_S(T)$)(left axis) and inverse initial susceptibility ($\chi_0^{-1}(T)$) (right axis) of (a) $Fe_2CrAl$ (b) $Fe_{1.75}Mn_{0.25}CrAl$ (c) $Fe_{1.5}Mn_{0.5}CrAl$ and (d) $Fe_{1.25}Mn_{0.75}CrAl$ obtained from high-field extrapolation of MAP (Fig.1 (a)-(d)). The values of critical exponents and $T_C$ are found from fitting of eqn. 1 and eqn. 2.

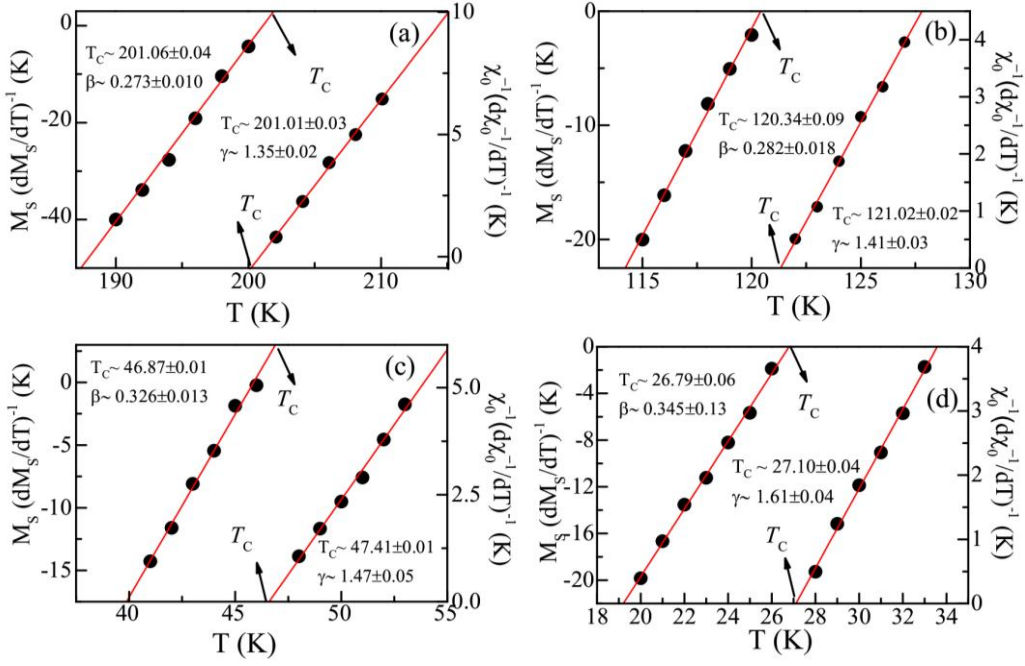

Figure 3 KF plot of spontaneous magnetization ($M_S$ (T)) and inverse susceptibility ($\chi_0^{-1}$ (T)) (right axis) of (a) $Fe_2CrAl$ (b) $Fe_{1.75}Mn_{0.25}CrAl$ (c) $Fe_{1.5}Mn_{0.5}CrAl$ and (d) $Fe_{1.25}Mn_{0.75}CrAl$. Solid red lines represent the linear fitting. The critical exponents and $T_C$ are estimated from the linear fit.

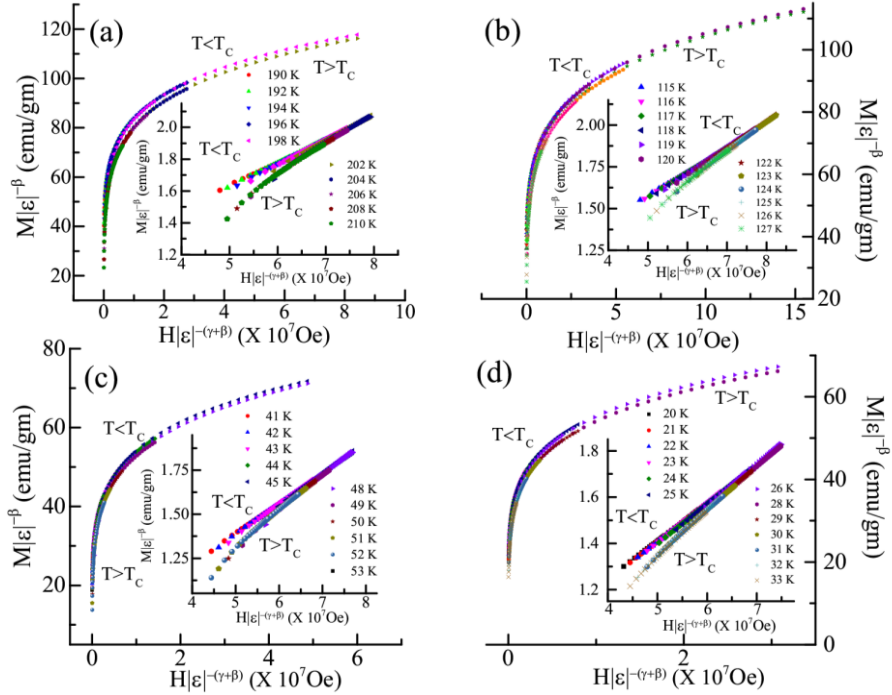

Figure 4 Renormalized magnetization as a function of renormalized field (eqn. 5) with critical exponents and $T_C$ from the table 1 for (a) $Fe_2CrAl$ (b) $Fe_{1.75}Mn_{0.25}CrAl$ (c) $Fe_{1.5}Mn_{0.5}CrAl$ and (d) $Fe_{1.25}Mn_{0.75}CrAl$. Plots show that all the data collapses into two different branches: one below $T_C$ and another above $T_C$. Insets: Same plot in the log-log scale.

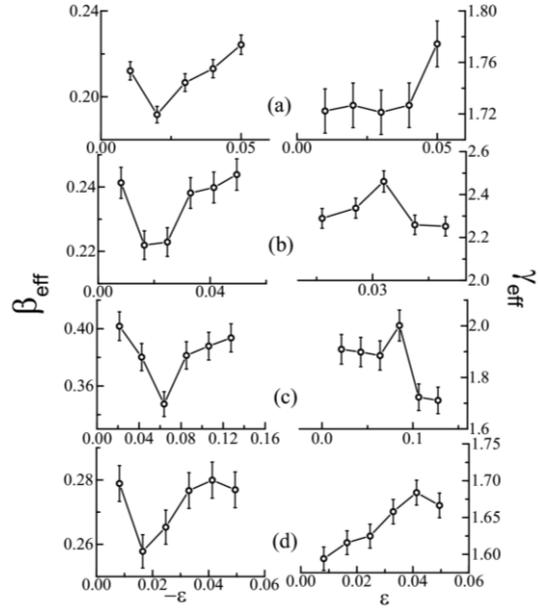

Figure 5 Effective critical exponents $\beta_{eff}$ below $T_C$ (left panel) and $\gamma_{eff}$ above $T_C$ (right panel) are plotted as a function of reduced temperature $\varepsilon$ for ((a) $Fe_2CrAl$ (b) $Fe_{1.75}Mn_{0.25}CrAl$ (c) $Fe_{1.5}Mn_{0.5}CrAl$ and (d) $Fe_{1.25}Mn_{0.75}CrAl$.

Supplementary material for

# Unconventional critical behaviour in weak ferromagnets Fe$_{2-x}$Mn$_x$CrAl (0≤x<1)

Kavita Yadav, Dheeraj Ranaut, and K. Mukherjee

School of Basic Sciences, Indian Institute of Technology, Mandi, Himachal Pradesh-175005, India

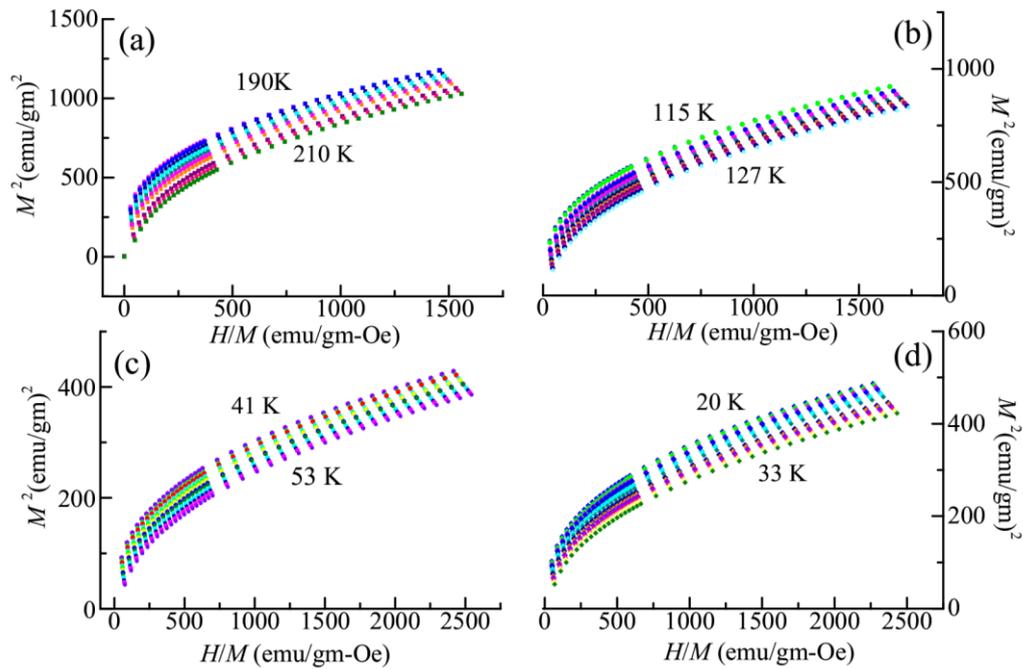

Fig. S1 Arrott plot of isotherms collected at different temperatures around $T_C$ for (a) Fe$_2$CrAl (b) Fe$_{1.75}$Mn$_{0.25}$CrAl (c) Fe$_{1.5}$Mn$_{0.5}$CrAl and (d) Fe$_{1.25}$Mn$_{0.75}$CrAl. Only few isotherms are shown for clarity.

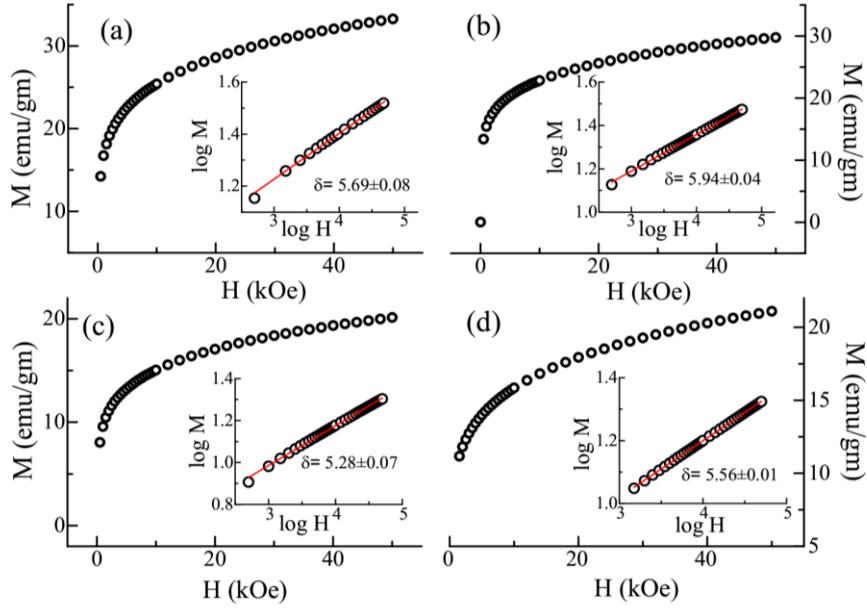

Fig. S2 M vs H plot at $T_C$ for (a) $Fe_2CrAl$ (b) $Fe_{1.75}Mn_{0.25}CrAl$ (c) $Fe_{1.5}Mn_{0.5}CrAl$ and (d) $Fe_{1.25}Mn_{0.75}CrAl$. Insets: Same plot in the log-log scale. Solid red lines represent the linear fitting.

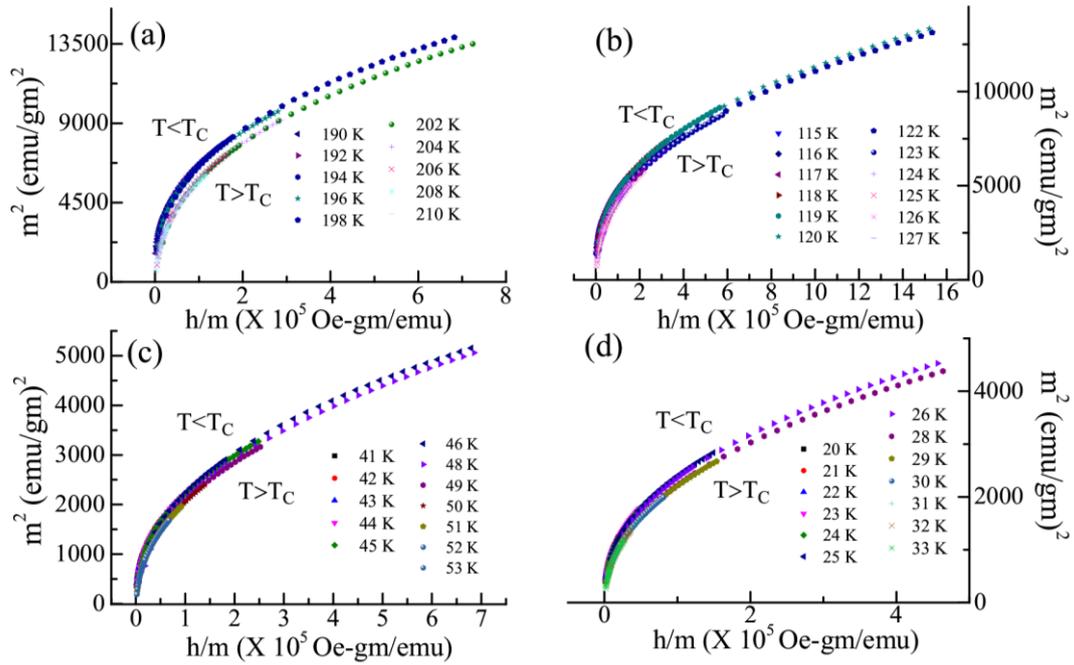

Fig. S3 Renormalized magnetization and field plotted in form of $m^2$ vs $h/m$ for (a) $Fe_2CrAl$ (b) $Fe_{1.75}Mn_{0.25}CrAl$ (c) $Fe_{1.5}Mn_{0.5}CrAl$ and (d) $Fe_{1.25}Mn_{0.75}CrAl$. The plot show that all the data fall into two distinct branches: one above $T_C$ and one below $T_C$.